\documentclass[twocolumn,prd,superscriptaddress,showpacs,floatfix,%
preprintnumbers,nofootinbib,showkeys]{revtex4}  

\usepackage{amsmath,amsfonts}
\usepackage{graphicx}

\def\bar {\overline}

\def\be {\begin{equation}}
\def\ee {\end{equation}}
\def\beq {\begin{equation}}
\def\eeq {\end{equation}}

\def\bea {\begin{eqnarray}}
\def\eea {\end{eqnarray}}

\def\bsbsbar{B_s$--$\bar{B}_s}

\def\dgs{\Delta \Gamma_s}

\def\beq{\begin{equation}}
\def\eeq{\end{equation}}
\def\barr{\begin{eqnarray}}
\def\earr{\end{eqnarray}}

\def\opcit(#1){ {\em op. cit.}, #1}
\def\etal {\em et al.}

\def\issue(#1,#2,#3){#1, #2 (#3)} 

\def\APP(#1,#2,#3){Acta Phys.\ Polon.\ \issue(#1,#2,#3)}
\def\ARNPS(#1,#2,#3){Ann.\ Rev.\ Nucl.\ Part.\ Sci.\ \issue(#1,#2,#3)}
\def\CPC(#1,#2,#3){Comp.\ Phys.\ Comm.\ \issue(#1,#2,#3)}
\def\CIP(#1,#2,#3){Comput.\ Phys.\ \issue(#1,#2,#3)}
\def\EPJC(#1,#2,#3){Eur.\ Phys.\ J.\ C\ \issue(#1,#2,#3)}
\def\EPJD(#1,#2,#3){Eur.\ Phys.\ J. Direct\ C\ \issue(#1,#2,#3)}
\def\IEEETNS(#1,#2,#3){IEEE Trans.\ Nucl.\ Sci.\ \issue(#1,#2,#3)}
\def\IJMP(#1,#2,#3){Int.\ J.\ Mod.\ Phys. \issue(#1,#2,#3)}
\def\JHEP(#1,#2,#3){J.\ High Energy Physics \issue(#1,#2,#3)}
\def\JPG(#1,#2,#3){J.\ Phys.\ G \issue(#1,#2,#3)}
\def\MPL(#1,#2,#3){Mod.\ Phys.\ Lett.\ \issue(#1,#2,#3)}
\def\NP(#1,#2,#3){Nucl.\ Phys.\ \issue(#1,#2,#3)}
\def\NIM(#1,#2,#3){Nucl.\ Instrum.\ Meth.\ \issue(#1,#2,#3)}
\def\PL(#1,#2,#3){Phys.\ Lett.\ \issue(#1,#2,#3)}
\def\PRD(#1,#2,#3){Phys.\ Rev.\ D \issue(#1,#2,#3)}
\def\PRL(#1,#2,#3){Phys.\ Rev.\ Lett.\ \issue(#1,#2,#3)}
\def\SJNP(#1,#2,#3){Sov.\ J. Nucl.\ Phys.\ \issue(#1,#2,#3)}
\def\ZPC(#1,#2,#3){Zeit.\ Phys.\ C \issue(#1,#2,#3)}

\begin{document}

\title{Enhanced $\bsbsbar$ lifetime difference 
and anomalous like-sign dimuon charge asymmetry
from new physics in $B_s \to \tau^+ \tau^-$}

\author{Amol Dighe}
\affiliation{Tata Institute of Fundamental Research, Homi Bhabha Road, 
Colaba, Mumbai 400005, India}
\author{Anirban Kundu}
\affiliation{Department of Physics, University of Calcutta, 92 Acharya 
Prafulla Chandra Road, Kolkata 700009, India}
\author{Soumitra Nandi}
\affiliation{Dipartimento di Fisica
Teorica, Univ. di Torino and INFN, Sezione di Torino, I-10125 Torino, Italy}

\begin{abstract} 
New physics models that increase the decay rate of $B_s \to \tau^+ \tau^-$
contribute to the absorptive part of $\bsbsbar$ mixing, and may enhance 
$\dgs$ all the way up to its current experimental bound.
In particular, the model with a scalar leptoquark 
can lead to a significant violation of the expectation
$\dgs \leq \dgs$ (SM).
It can even allow regions in the $\Delta \Gamma_s$--$\beta_s$ 
parameter space that are close to the best fit 
obtained by CDF and D\O~through $B_s \to J/\psi \phi$.
In addition, it can help explain the anomalous like-sign
dimuon charge asymmetry observed recently by D\O.
A measurement of $BR(B_s \to \tau^+ \tau^-)$ is thus crucial for
a better understanding of new physics involved in $\bsbsbar$
mixing.
\end{abstract}

\date{\today}




\maketitle

\section{Introduction}
\label{intro}

In the standard model (SM), the Cabbibo-Kobayashi-Maskawa (CKM) 
mixing matrix is the only source of charge-parity (CP) violation.
The data from the decays of $K$, $D$ and $B$ mesons have so
far been consistent with this paradigm, however the flavor changing
neutral current (FCNC) processes involving $b \to s$ transitions
are expected to be sensitive to many sources of new physics (NP)
\cite{bf-review}. This is why the $B_s$ meson is one of the most
important and interesting portals for indirect detection
of such NP models.

\begin{figure}
\includegraphics[width= 0.85 \linewidth]{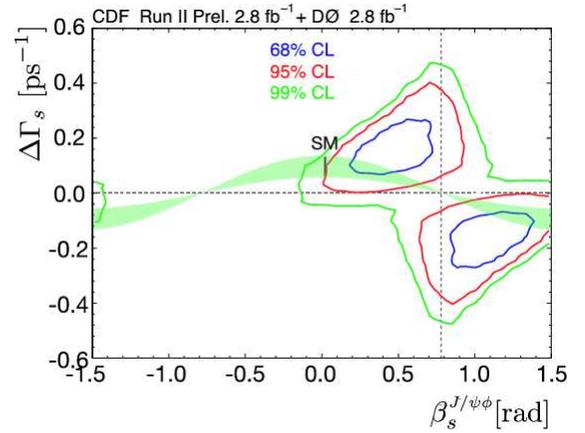}
\caption{The combined experimental constraints by CDF and 
D\O~ through $B_s \to J/\psi \phi$. 
Blue, red and green contours (from inner to outer)
correspond to the 68\%, 95\% and 99\% C.L. regions.
The sinusoidal green band corresponds to the relation 
$\dgs \approx \dgs^{\rm SM} \cos\phi_s$, valid when NP does not
contribute to $\Gamma_{12s}$.
The figure is taken from \cite{pubn}.
\label{fig:fitonly}}
\end{figure}

In this paper we shall concentrate on the oscillation parameters in the 
$\bsbsbar$ system. The average decay width
$
\bar{\Gamma}_s \equiv (\Gamma_{sH}+\Gamma_{sL})/2 = 
\left(0.679^{+0.013}_{-0.011} \right)\mbox{ps}^{-1}
$
and the mass difference 
$\Delta M_s \equiv M_{sH} - M_{sL} = \left(17.77 \pm 0.10 \pm 0.07
\right)\mbox{ps}^{-1}
$
have already been measured to an accuracy of better than $\sim 2\%$ 
\cite{pdg,deltams,utfit} and play an important role in constraining 
any new physics.
Here the labels $L$ and $H$ stand respectively for the light and heavy 
mass eigenstates in the neutral $B_s$ system.
The decay width difference $\Delta\Gamma_s \equiv \Gamma_{sL}-\Gamma_{sH}$ 
and the $\bsbsbar$ mixing phase are relatively less certain.
The SM predictions for these quantities are \cite{lenz-nierste}
\bea
\dgs^{\rm SM} & = & (0.096 \pm 0.039) \mbox{ ps}^{-1} \; ,\\
 \beta_s^{J/\psi \phi {(\rm SM})} & = & {\rm Arg}
\left(- \frac{V_{cb} V_{cs}^*}{V_{tb} V_{ts}^*} \right) 
\approx 0.019 \pm 0.001 ,
\eea
where $2\beta_s^{J/\psi \phi}$ is the mixing phase
relevant for $B_s \to J/\psi \phi$ decay.
The recent CDF and D\O~measurements \cite{cdfd0,pubn}, using the angular
analysis in $B_s \to J/\psi \phi$ decay \cite{ddlr,ddf}, 
give \cite{hfag}
\bea 
\dgs & = & \pm (0.154 ^{+0.054}_{-0.070}) \mbox{ ps}^{-1}\;, \\
\beta_s^{J/\psi \phi} & = & (0.39^{+0.18}_{-0.14})\cup (1.18^{+0.14}_
{-0.18}) \; ,
\eea
where the second set in the last line is just the complement of $\pi/2$ 
for the first set. This reflects the ambiguity in the determination of
$\beta_s^{J/\psi \phi}$.
Note that the sign of $\dgs$ is undetermined. The positive and
negative signs correspond, respectively, to the two disconnected 
regions in the allowed parameter space for $\beta_s^{J/\psi \phi}$.
Alternative ways of removing this sign ambiguity have been 
suggested in \cite{nandi_nierste}.
The correlated constraints are shown in Fig.~\ref{fig:fitonly}.
The SM prediction for $(\dgs,\beta_s^{J/\psi\phi})$ is excluded by the data 
to 90\% C.L.. Hence the exploration of new physics effects on these
quantities becomes imperative.

While many new physics models can affect $\beta_s^{J/\psi\phi}$ and make its
value anywhere in its conventional allowed range $[-\pi/2, \pi/2]$,
the ability of new physics to influence $\dgs$ is rather limited.
Indeed, the width difference is 
\be
\dgs = 2 |\Gamma_{12s}| \cos\phi_s \; ,
\label{cosphi}
\ee
where 
$\phi_s \equiv {\rm Arg}(-M_{12s}/\Gamma_{12s})$.
Here $M_{12s}$ and $\Gamma_{12s}$ are the dispersive and
absorptive parts, respectively, of the $\bsbsbar$ mixing amplitude.
In the SM \cite{lenz-nierste}
\beq
\phi_s = 0.0041 \pm 0.0007 \; ,
\label{phis-def}
\eeq
and hence $\dgs^{\rm SM} \approx 2 |\Gamma_{12s}|$.
The class of NP models which do not affect $\Gamma_{12s}$
then satisfy $\dgs \leq \dgs^{\rm SM}$ \cite{grossman}.
These include the minimal flavour violating models \cite{mfv}
where the bases in the quark flavor space are the same as that
in the SM,
as well as models where the mixing box diagram contains
only heavy degrees of freedom.
The predictions of these models for $(\dgs, \beta_s^{J/\psi \phi})$ will 
then be restricted to the sinusoidal band shown in Fig.~\ref{fig:fitonly}.
Note that only a small part of this band is within the 68\% C.L. 
region, so that NP of this type will be unable to 
account for the measurements if the errors decrease with
the best fit values staying unchanged.

However, there are well-motivated models where the $\bsbsbar$
mixing box diagram contains two light degrees of freedom, 
resulting in an absorptive amplitude. 
Given the current strong constraints on the $B_s$ decays to
hadrons, $e^+ e^-$ and $\mu^+ \mu^-$ \cite{pdg},
the only candidate for the intermediate light particle is $\tau$.
In an earlier publication \cite{dkn1}, we had implemented
this idea with two examples:
(i) the model with a scalar leptoquark, and 
(ii) R-parity violating supersymmetry.
These models can have {\em flavor dependent} couplings of 
a light particle with a heavy new particle -- in particular, 
$\tau$ can couple with the leptoquark or squark -- 
and hence can contribute to $\Gamma_{12s}$.
A significant enhancement of $\dgs$
was shown to be possible in the former 
model \cite{dkn1}. 
In this paper we shall investigate the effect of the 
leptoquark on the correlation between $\dgs$ and $\beta_s^{J/\psi\phi}$,
keeping in mind that any such new physics will also significantly
affect the decay rate $B_s \to \tau^+ \tau^-$.

Recently, the D\O~ collaboration has  claimed evidence
for an anomalous like-sign dimuon charge asymmetry \cite{d0-6.1}
\beq
A_{\rm sl}^b = - 0.00957 \pm 0.00251 \pm 0.00146 \; .
\eeq
CDF has also measured the same quantity using 1.6 fb$^{-1}$ of data and found
$A^b_{\rm sl} = (8.0\pm 9.0\pm 6.8)\times 10^{-3}$ \cite{cdf-1.6}.
Combining these two, one gets
\beq
A_{\rm sl}^b = -(8.5\pm 2.8)\times 10^{-3}\,,
\label{aslb-avg}
\eeq
which differs from the SM prediction
\beq
A_{\rm sl}^{b {\rm (SM)}} = -0.00023 ^{+0.00005}_{-0.00006}
\eeq
by about 3$\sigma$.
Such an asymmetry can be used as a probe of the flavor structure of
new physics \cite{randall-su}.
It turns out that the same new physics that enhances $\Delta\Gamma_s$
can also help in explaining this anomaly.
We shall elaborate on this in the latter part of this paper.

\section{New physics in $B_s\to \tau^+\tau^-$}
\label{formalism}

Leptoquarks (LQ) are color-triplet objects that couple to
quarks and leptons. They occur generically in
GUTs \cite{lq-gut}, composite models \cite{lq-composite},
and superstring-inspired $E_6$ models \cite{lq-e6}.
Model-independent constraints on their properties are
available \cite{davidson-zpc61}, and the prospects of
their discovery at the LHC have also been studied \cite{mitsou}.

The direct production limits depend on the LQ model, as well as the
SM fermions these LQs can couple to.
The bounds on the second and third generation leptoquarks are,
respectively, $M_{\rm LQ} > 316, 245$ GeV, when they are pair produced
\cite{pdg,d0-leptoquark}.
A third generation scalar leptoquark decaying only into a $b$-quark and
a $\tau$ lepton has a mass bound of 210 GeV
\cite{d0:0806.3527}.
We shall conservatively take $M_{\rm LQ}=250$ GeV in this analysis.
However our results hold even with much higher $M_{\rm LQ}$, by
appropriately scaling the coupling $|h_{\rm LQ}|$ as shall be seen later.

We shall restrict ourselves to scalar leptoquarks that are
singlets under the SU(2)$_L$ gauge group of the SM.
This is because vector or 
most of the SU(2)$_L$ nonsinglet leptoquarks tend to
couple directly to neutrinos, hence we expect that their couplings 
are tightly constrained from the neutrino mass and mixing data.
This makes any significant effect on the $\bsbsbar$ system
unlikely.

The relevant interaction term for a scalar SU(2)$_L$
singlet leptoquark is
of the form
\be
{\cal L}_{\rm LQ} = \lambda_{ij} \bar{d^c}_{jR}
e_{iR}{\cal S}_0 + {\rm h.c.}  \; ,
\label{lqlag}
\ee
where $d_R$ and $e_R$ stand for the right-handed down-type
quarks and right-handed charged leptons, respectively, and
$i,j$ are generation indices that run from 1 to 3.
The couplings $\lambda_{ij}$ can in general be complex,
and some of them may vanish depending on any flavor
symmetries involved. 
We take the LQ couplings in the quark mass basis. This is the most
economical choice given the fact that we do not know the rotation
matrix for the right-chiral down-type quark fields.
One can also have an SU(2)$_L$ doublet
leptoquark, whose interaction is of the form
\be
{\cal L}_{\rm LQ} = \lambda_{ij} \bar{q}_{jL} i\sigma_2
e_{iR}{\cal S}_\frac12 + {\rm h.c.}  \; ,
\label{lqlag2}
\ee
which gives almost identical results.

When $\lambda_{32}$ and $\lambda_{33}$ are nonzero,
the interaction in eq.~(\ref{lqlag}) generates
an effective four-fermion
$(S+P)\otimes (S+P)$ interaction leading to $b\to s\tau^+\tau^-$.
This will contribute to $\bsbsbar$ mixing
(with $\tau$ and ${\cal S}_0$ flowing inside the box),
to the leptonic decay $B_s\to \tau^+\tau^-$, and to the semileptonic
decays $B \to X_s \tau^+\tau^-$. 
The relevant quantity here is the coupling product
\beq
h_{\rm LQ} (b \to s \tau^+ \tau^-) \equiv
\lambda_{32}^\ast \lambda_{33} \; .
\eeq

One may get a tight constraint on $|h_{\rm LQ}|$ from
$B_s \to \tau^+ \tau^-$.
One expects the lifetimes of $B_d$ and $B_s$ to be the same in the Standard
Model: $\tau_{B_s}/\tau_{B_d} = 1.00\pm 0.01$ \cite{pdg}. 
This is certainly true if we assume spectator dominance: the decays which
do not have a spectator quark contribute negligibly in the total decay
width. Experimentally, 
$\Gamma_s/\Gamma_d - 1 = (3.6\pm 1.8)\%$ \cite{hfag}.
Thus, the branching ratio ${\cal B}\equiv {\rm BR}(B_s\to\tau^+\tau^-)$ 
can be as large as 6--7\%.
Considering the deviations from the naive spectator model, which is expected
to be small for the $B_d$-$B_s$ system, one may conservatively put the upper 
bound for ${\cal B}$ at 10\%.
The value of ${\cal B}$ is only
${\cal O}(10^{-8})$ in the SM. 
This decay has not been observed, nor is a direct measurement of 
an upper bound on its branching ratio available. 
A similar estimate of ${\cal B}\sim 5\%$ is available in
\cite{gln-tau}.
If indeed $|h_{\rm LQ}|$ is large enough to cause such a significant
enhancement in ${\cal B}$, it is related to ${\cal B}$
directly through
\begin{eqnarray}
&& {\cal B} \approx 
\frac{|h_{\rm LQ}|^2}{ 128 \pi M_{\rm LQ}^4}
 \frac{f_{B_s}^2 M_{B_s}^3}{\bar\Gamma_{s}} \frac{m_\tau^2}
{M_{B_s}^2} \sqrt{1-4\frac{m_{\tau}^2}{M_{B_s}^2}} \; \nonumber \\
 &&\approx 9.5 \% \, \left(\frac{|h_{\rm LQ}|}{0.3}\right)^2\, 
\left(\frac{250 {\rm ~GeV}}{M_{\rm LQ}}\right)^4\,
\left(\frac{f_{B_s}}{0.250 {\rm  ~GeV}}\right), \phantom{spa}
\end{eqnarray}
where $f_{B_s}$ is the $B_s$ decay constant.
It can be seen that for $M_{\rm LQ}= 250$ GeV,
${\cal B} \approx 10\%$ can accommodate
$|h_{\rm LQ}| \approx 0.3$.

We shall show in the next sections that the values of $|h_{\rm LQ}|$ allowed
by the above analysis can cause significant changes in the 
values of $\dgs$ and $\beta_s^{J/\psi \phi}$,
and can also enhance $A_{sl}^b$ by a sizeable amount.

\section{New Physics in $\dgs$ and $\beta_s^{J/\psi \phi}$}
\label{sec:Jpsiphi}       

In the presence of NP contribution, the expressions for the dispersive 
and absorptive parts of $\bsbsbar$ mixing can be written as
\bea
M_{12s} & = & M_{12s}^{\rm SM} + M_{12s}^{\rm LQ} 
= M_{12s}^{\rm SM} \, R_M \, e^{i\phi_M} \; , 
\label{M12-net} \\
\Gamma_{12s} & = & \Gamma_{12s}^{\rm SM} + \Gamma_{12s}^{\rm LQ} 
= \Gamma_{12s}^{\rm SM} \, R_\Gamma \, e^{i \phi_\Gamma} \; . 
\label{G12-net}
\eea
The standard model contributions, to leading order (LO) in $1/m_b$ and 
$\alpha_s(m_b)$, are given by \cite{hagelin,gamma12-lo}
\bea
M_{12s}^{\rm SM} & = &  (V_{tb} V_{ts}^*)^2 
\frac{G_F^2}{12\pi^2} 
\chi_{B_s} \hat{\eta}_{B_s} M_W^2 S_0(x_t) \; , 
\label{M12sm} \\
\Gamma_{12s}^{\rm SM} & = &   - 
[ (V_{cb} V_{cs}^*)^2 \Gamma^{cc} +
(V_{ub} V_{us}^*)^2 \Gamma^{uu} \nonumber \\
& &  \phantom{spacespace} + 2(V_{cb} V_{cs}^* V_{ub} V_{us}^*) \Gamma^{cu}]\; ,
\label{G12sm}
\eea
where $\chi_{B_s} \equiv M_{B_s} B_{B_s} f_{B_s}^2$,
and $\Gamma^{ij}$, the absorptive parts of the box diagrams 
(without the CKM factors) with quarks $i$ and $j$ flowing inside 
the loop, are given in \cite{lenz-nierste}.
The short distance behavior is contained in 
$\hat{\eta}_{B_q}$, which incorporates the QCD corrections,
and in the Inami-Lim function $S_0(x_t)$.
The value of $\Gamma_{12s}^{\rm SM}$ has been calculated 
up to ${\cal O}(1/m_b^2)$ in \cite{badin}, wherein some NP contributions 
to $\Delta\Gamma_s$ have also been studied.

The leading order leptoquark contributions to the
above quantities are \cite{dkn1}
\bea
M_{12s}^{\rm LQ} & = & \frac{h_{\rm LQ}^2}{384\pi^2 M_{\rm LQ}^2} 
\chi_{B_s} \hat{\eta}_{B_s} \tilde{S_0}(x_\tau) \; ,  
\label{M12lq} \\
\Gamma_{12s}^{\rm LQ(0)} & = & - 
\frac{h_{\rm LQ}^2}{256 \pi M_{\rm LQ}^4} 
\chi_{B_s}  m_b^2 F(\tau) \; ,
\label{G12lq}
\eea
where $\tilde{S_0}(x_\tau)$ is another Inami-Lim function,
and the phase space factor is $F(\tau) = 0.64$.
The details of the calculation may be found in \cite{dkn1}.

\begin{figure}[t]
\centerline{\includegraphics[width= 0.85 \linewidth]{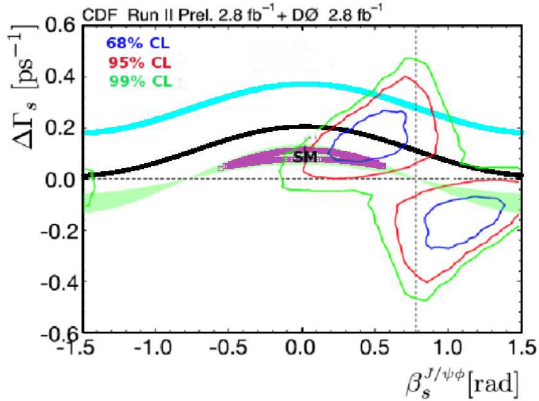}}
\caption{The predictions of $(\dgs,\beta_s^{J/\psi\phi})$ 
within the scalar leptoquark model, overlayed on the 
combined experimental constraints by CDF and D\O~ through
$B_s \to J/\psi \phi$ (Fig.~\ref{fig:fitonly}). 
Magenta (dark gray), black and aqua (light gray) bands correspond to 
$|h_{LQ}| = 0.07,\, 0.17$ and $0.27$, respectively,
with $M_{\rm LQ} = 250$ GeV.
\label{fitwithpoints}}
\end{figure}

While the next to leading order QCD corrections and the $1/m_b$ corrections 
do not affect $M_{12s}^{\rm SM}$ significantly, they
modify $\Gamma_{12s}^{\rm SM}$ by $\sim 30\%$ from its LO value \cite{nlo-dgs}.
The QCD corrections are expected to be different for SM
and LQ operators since the mediating heavy particle for the latter
case is a color triplet. The $1/m_b$ corrections are also expected
to differ since the light degrees of freedom that flow inside the
mixing box are different too. While it is desirable to have an
idea of these corrections, 
since we are only showing typical results from
allowed leptoquark parameters, such corrections can be absorbed 
by just changing the value of $M_{\rm LQ}$ and the phase of $h_{\rm LQ}$.
Therefore in our numerical analysis, we use the SM predictions 
for $\Gamma_{12s}^{\rm SM}$ \cite{lenz-nierste} 
that include the NLO QCD and $1/m_b$ corrections,
however for $\Gamma_{12s}^{\rm LQ}$ we only use the leading order contribution. 
For the sake of clarity, while calculating the combined SM and LQ
contribution to $\Gamma_{12s}$, we use only the central value of the 
SM prediction. Including the 30\% error in the SM prediction will 
widen the bands for our results shown in Fig.~\ref{fitwithpoints}.

In the presence of leptoquarks, eqs.~(\ref{cosphi},\ref{M12-net},\ref{G12-net})
lead us to write the width difference as
\bea
\dgs & = & 2 |\Gamma_{12s}^{\rm SM}| \, R_\Gamma \, 
\cos(\phi_M - \phi_\Gamma - 2\beta_s^{\rm SM}) \nonumber \\
& \approx & \dgs^{\rm SM} \, R_\Gamma \, \cos(\phi_M - \phi_\Gamma) \; ,
\eea
where the approximation uses $\beta_s^{\rm SM} \approx 0$.
The allowed values of $h_{\rm LQ}$ permit
$R_\Gamma \cos(\phi_M - \phi_\Gamma) > 1$, so that 
the value of $\dgs$ can be enhanced in this model. 
Fig.~\ref{fitwithpoints} shows that the enhancement can be 
even up to $\dgs \approx 0.4$ ps$^{-1}$ for $|h_{LQ}| \approx 0.3$.

The decay $B_s \to J/\psi \phi$ exhibits CP violation through
the interference of mixing and decay.
The CP violating phase measured through the time dependent
angular distribution of this decay is
\beq
\beta_s^{J/\psi \phi} \approx \frac{1}{2}
{\rm Arg}\left( - \frac{(V_{cb} V_{cs}^*)^2}{M_{12s}} \right) 
= \beta_s^{J/\psi \phi {\rm (SM)}} - \frac{\phi_M}{2} \; ,
\label{betaphi}
\eeq
where we have used the approximation $|\Gamma_{12s}| \ll |M_{12s}|$. 
Clearly at low values of $|h_{\rm LQ}|$, the allowed range of $\phi_M$
will be restricted to be near zero, and hence $\beta_s^{J/\psi \phi}$ will be
close to its SM value, which itself is close to zero. For higher
$|h_{\rm LQ}|$, however, the value of $\phi_M$ can be anything, 
and hence $\beta_s^{J/\psi \phi}$
can be anywhere in its conventional range $[-\pi/2, \pi/2]$. 
This is illustrated in Fig.~\ref{fitwithpoints}.

Figure~\ref{fitwithpoints} overlays our predictions with the leptoquark
model in the $\dgs$--$\beta_s^{J/\psi \phi}$ plane on the results of 
the combined analysis of CDF and D\O.~
Clearly, the additional leptoquark contribution not only can enhance
$\dgs$ and $\beta_s$, but also can allow us to be well within the
68\% C.L. region of the current best fit.

\section{New Physics in $A_{\rm sl}^b$}
\label{dimuon}

The like-sign dimuon charge asymmetry $A_{\rm sl}^b$ 
measured by D\O~ \cite{d0-6.1} and CDF \cite{cdf-1.6} is
related to the semileptonic decay asymmetries $a_{\rm sl}^d$ and
$a_{\rm sl}^s$ in the $B_d$ and $B_s$ sectors, respectively, 
through \cite{d0-6.1}
\beq
A_{\rm sl}^b = (0.506\pm0.043) \, a_{\rm sl}^d
+(0.494\pm 0.043)\, a_{\rm sl}^s \;.
\eeq
The coefficients here are valid even in the presence of NP. 
The average $A_{\rm sl}^b$ from eq.~(\ref{aslb-avg}),
and the current experimental constraints of
$a_{\rm sl}^d = -0.0047 \pm 0.0046$ \cite{hfag}, yield
\beq
a_{\rm sl}^s = -0.012 \pm 0.007 \; ,
\eeq
which is almost $2\sigma$ away from the SM prediction \cite{lenz-nierste}
\beq
a_{\rm sl}^{s {\rm (SM)}} = (2.1 \pm 0.6) \times 10^{-5} \; .
\eeq
This quantity is directly related to $\dgs$ and the $\bsbsbar$
mixing phase via
\beq
a_{\rm sl}^s = \frac{\dgs}{\Delta M_s} \tan\phi_s^{\rm sl} =
- \frac{\dgs}{\Delta M_s} \tan 2\beta_s^{\rm sl} \; ,
\eeq
where $\phi_s^{\rm sl} \equiv {\rm Arg}(-M_{12s}/\Gamma_{12s})=\phi_s$
and we have defined $\beta_s^{\rm sl}$ such that 
$\phi_s^{\rm sl} = - 2 \beta_s^{\rm sl}$.
From eq.~(\ref{phis-def}), we have
\beq
\beta_s^{\rm sl} = -0.0020 \pm 0.0003 \; .
\eeq

In the presence of NP that affects $\Gamma_{12s}$, 
eqs.~(\ref{M12-net}) and (\ref{G12-net}) yield the relation
\beq
\beta_s^{\rm sl} = \frac{1}{2}
{\rm Arg}\left( -\frac{\Gamma_{12s}}{M_{12s}} \right)
 = \beta_s^{\rm sl (SM)} 
- \frac{\phi_M}{2} + \frac{\phi_\Gamma}{2} \;.
\label{betasl}
\eeq
Since $\beta_s^{J/\psi \phi {\rm (SM)}} \approx 0 \approx 
\beta_s^{\rm sl (SM)}$, eqs.~(\ref{betaphi}) and (\ref{betasl})
clearly show that 
$\beta_s^{\rm sl}$ is in general different from $\beta_s^{J/\psi \phi}$.
Note that when NP does not affect $\Gamma_{12s}$, the value of
$\phi_\Gamma$ vanishes and only then can one say
$\beta_s^{\rm sl} \approx \beta_s^{J/\psi\phi}$.
Therefore,
it is not recommended to superimpose the parameter spaces of
$(\dgs, \beta_s^{\rm sl})$ and $(\dgs, \beta_s^{J/\psi \phi})$.

\begin{figure}[t]
\centerline{
\includegraphics[width= 0.80 \linewidth]{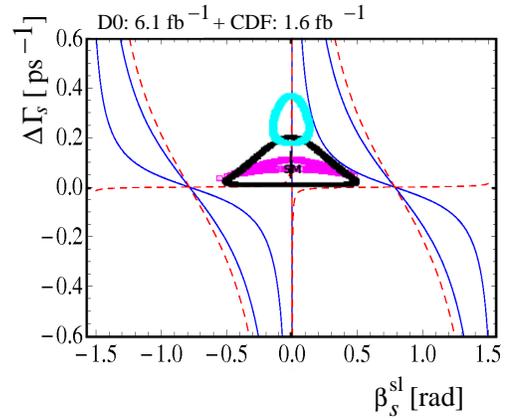}}
\caption{The predictions for $(\dgs,\beta_s^{\rm sl})$ 
within the scalar leptoquark model, overlayed on the
68\% (continuous blue) and 95\% (dashed red) C.L. contours 
for the combined
D\O~ and CDF measurements of $a_{\rm sl}^s$.
Magenta (dark gray), black and aqua (light gray) bands correspond to 
$|h_{LQ}| = 0.07,\, 0.17$ and $0.27$, respectively,
with $M_{\rm LQ} = 250$ GeV. 
\label{fig:sl}}
\end{figure}

In Fig.~\ref{fig:sl}, we show the constraints in the
$(\dgs, \beta_s^{\rm sl})$ parameter space
coming from the $A_{\rm sl}^b$
(consequently, $a_{\rm sl}^s$) measurement in \cite{d0-6.1},
and $(\dgs, \beta_s^{\rm sl})$ predictions 
at some allowed $|h_{LQ}|$ values.
It shows that the leptoquark contribution can give rise
to $a_{\rm sl}^s$ values well within the 95\% C.L. region of the 
experimental data.
Note that the predictions shown in Figs.~\ref{fitwithpoints}
and \ref{fig:sl} correspond to the same set of NP parameters.
This again illustrates the need to clearly differentiate between
$\beta_s^{\rm sl}$ and $\beta_s^{J/\psi \phi}$. 

\section{Summary and conclusions}
\label{summary}

The model with a scalar leptoquark, presented in this paper, belongs to
the special class of NP models that affect the absorptive part
$\Gamma_{12s}$ of $\bsbsbar$ mixing. It can therefore evade the
relation $\dgs < \dgs^{SM}$ and can give enhanced values of the
lifetime difference in $\bsbsbar$ system. The enhancement in
$\dgs$ also corresponds to an enhancement in the branching ratio
$BR(B_s \to \tau^+ \tau^-)$. 

Recent measurements of  $\dgs$ and $\beta_s^{J/\psi \phi}$ by the CDF 
and D\O~collaborations exclude the SM prediction to 90\% C.L..
We illustrate with the example of the scalar leptoquark model that
$\dgs$ as large as 0.4 ps$^{-1}$ may be achieved, and values in
the $(\dgs, \beta_s^{J/\psi \phi})$ parameter space close
to the best fit from these measurements can be obtained. 
Indeed, if future experiments decrease the errors on these quantities while
keeping the best fit values at their current positions, only models
belonging to this class will be able to explain the
deviation from the SM.

The explanation of anomalous like-sign dimuon charge asymmetry recently 
observed at D\O~ is also facilitated by this class of models,
since these models give rise to large $\dgs$ as well as large $\beta_s^{\rm sl}$
simultaneously. We point out that these models in general
imply that $\beta_s^{\rm sl} \neq \beta_s^{J/\psi\phi}$, so one has to be 
careful when including NP in the analysis.
Also, note that this mechanism affects $A_{\rm sl}^b$ through the
modification of $\dgs$ and $\phi_s$, without the need of an
explicit $b \to s \mu^+ \mu^-$ coupling.
This is a common feature of all models which have an absorptive part
in the $\bsbsbar$ mixing diagram.

In order to confirm the compatibility of
such models with the data, one needs further NLO calculations
of the predictions of these models, as well as a better measurement
of $B_s \to \tau^+ \tau^-$ branching ratio, which will be crucial
to constrain the leptoquark couplings. 
The $\tau$ from $B_s \to \tau^+ \tau^-$ may be expected to have enough
energy boost at the LHC to be detected. The $\tau$ polarization can
also be measured: the $\tau$'s coming from leptoquarks are expected to be
right handed. In addition, if we have an SU(2)$_L$ doublet leptoquark
${\cal S}_{\frac{1}{2}}$, this will also give rise to the FCNC top
decay $t \to c \tau^+ \tau^-$ at the level of 1\%, which will be
another probe of the new physics of this class.

\section*{Acknowledgements}

We thank B. Dobrescu and D. Hedin for pointing us to the latest bounds on 
leptoquark masses, and A. Lenz for his clarification about
various $\beta_s$ phases.
A.K. acknowledges CSIR, Government of India, and the DRS programme of
UGC, Government of India, for financial support. S.N. is 
supported in part by MIUR under contract  2008H8F9RA$\_$002
and by the EU's Marie-Curie Research Training Network under
contract MRTN-CT-2006-035505 `Tools and Precision Calculations for Physics
Discoveries at Colliders'.


\end{document}